\documentclass[aps,prl,twocolumn,superscriptaddress]{revtex4}
\usepackage[colorlinks=true,urlcolor=blue,citecolor=blue,linkcolor=blue]{hyperref}
\usepackage{graphicx}
\usepackage{latexsym}
\usepackage{stmaryrd}
\usepackage{amsmath}
\usepackage{amssymb}
\usepackage{epstopdf}
\usepackage{slashed}
\usepackage{mathrsfs}
\usepackage{bm}
\usepackage{verbatim}
\usepackage{hyperref}

\begin{document}
\makeatletter
\newcommand{\rmnum}[1]{\romannumeral #1}
\newcommand{\Rmnum}[1]{\expandafter\@slowromancap\romannumeral #1@}
\newcommand{\B}[1]{{\textcolor{blue}{#1}}}
\makeatother
	
\title{Chern Vector Protected Three-dimensional Quantized Hall Effect}
\author{Zhi-Qiang Zhang }
\affiliation{Interdisciplinary Center for Theoretical Physics and Information Sciences (ICTPIS), Fudan University, Shanghai 200433, China}
\affiliation{School of Physical Science and Technology, Soochow University, Suzhou 215006, China}
\author{Shu-Guang Cheng }
\affiliation{Department of Physics, Northwest University, Xi'an 710069, China}
\author{Hongfang Liu }
\affiliation{School of Physical Science and Technology, Soochow University, Suzhou 215006, China}
\author{Hailong Li }\email{hailongli@fudan.edu.cn}
\affiliation{Interdisciplinary Center for Theoretical Physics and Information Sciences (ICTPIS), Fudan University, Shanghai 200433, China}
\author{Hua Jiang}\email{jianghuaphy@fudan.edu.cn}
\affiliation{Interdisciplinary Center for Theoretical Physics and Information Sciences (ICTPIS), Fudan University, Shanghai 200433, China}
\author{X. C. Xie}
\affiliation{Interdisciplinary Center for Theoretical Physics and Information Sciences (ICTPIS), Fudan University, Shanghai 200433, China}
\affiliation{International Center for Quantum Materials, School of Physics, Peking University, Beijing 100871, China}
\affiliation{Hefei National Laboratory, Hefei 230088, China}
\date{\today}
	
\begin{abstract}
Recently, Chern vector with arbitrary formula $\textbf{C}\!=\!(\mathcal{C}_{yz},\mathcal{C}_{xz},\mathcal{C}_{xy})$ in three-dimensional systems has been experimentally realized [\B{Nature 609, 925 (2022)}]. 
Motivated by these progresses, we propose the Chern vector $\textbf{C}\!=\!(0,m,n)$-protected quantized Hall effect in three-dimensional systems.
By examining samples with Chern vector $\textbf{C}\!=\!(0,m,n)$ and dimensions $L_y$ and $L_z$ along the $y$- and $z$-directions, we demonstrate a topologically protected two-terminal response.
This response can be reformulated as the sum of the transmission coefficients along the $x$- and $y$-directions, given by $(mL_y\!+\!nL_z)$.
When applied to Hall bar setups, this topological mechanism gives rise to quantized Hall conductances, such as \(G_{xy}\) and \(G_{xz}\), which are expressed by $\pm(mL_y\!+\!nL_z)$.
These Hall conductances exhibit a clear dependency on sample dimensions, illuminating the intrinsic three-dimensional nature.
Finally, we propse potential candidates for experimental realization.
Our findings not only deepen the understanding of the topological nature of Chern vectors but also enlighten the exploration of their transport properties.
\end{abstract}
	
\maketitle

\begin{figure}[t]
\centering
\includegraphics[width=0.48\textwidth]{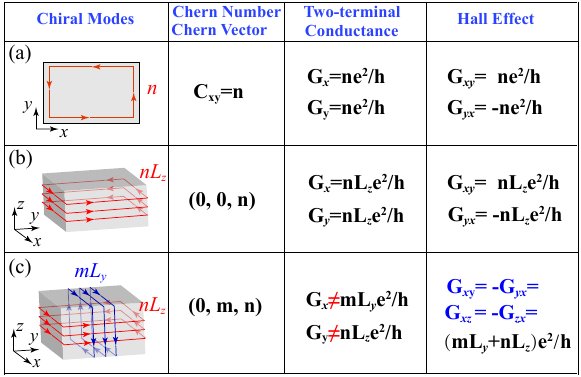}
	\caption{Comparing of the chiral modes, Chern number (vectors), conductances and Hall effects for (a) 2D samples with $\mathcal{C}_{xy}\!=\!n$; (b) $\textbf{C}\!=\!(0,0,n)$; (c) $\textbf{C}\!=\!(0,m,n)$. The sample sizes are $L_\alpha$ with $\alpha\!\in\! x,y,z$. $G_\alpha$ is the two-terminal conductance along $\alpha$ direction. We fix the Chern vector in the positive direction with $m\!\ge\!0$ and $n\!\ge\!0$.}
	\label{f1}
\end{figure}

\B{\textit{Introduction.---}}
The two-dimensional quantized Hall effect (QHE) \cite{zongshu1,zongshu2,shen1,QH1,QH2,QH3,QAH1,QAH2,QAH3,QAH6,QAH7,QAH8,QAH9,QAH10,QAH11,QAH12,Mn5,3DQHCDW,3DQHE1,3DQHE2} has long been a central topic in topological physics, establishing a key connection between the Chern number $\mathcal{C}_{xy}$, the chiral modes and the Hall conductance $G_{xy}\!=\!\mathcal{C}_{xy}e^2/h$ \cite{QH2,QH3,QAH1}, as shown in Fig. \ref{f1} (a).
Recently, interest has expanded to three dimensional (3D) systems where various Hall responses have been proposed and experimentally verified \cite{QAH11,QAH12,Mn5,3DQHCDW,3DQHE1,3DQHE2,3DQHE3,3DQHE4,3DQHE5,3DHE1,3DHE2,3DHE3,3DHE4,3DHE5}.
Significantly, the concept of the Chern number has been extended to 3D systems by introducing the Chern vector $\textbf{C}\!=\!(\mathcal{C}_{yz},\mathcal{C}_{xz},\mathcal{C}_{xy})$ \cite{Chernvector,Chernvector2,Chernvector3}, whose Chern number components are believed to govern both the spatial distribution of chiral modes and their transport properties.
For example, when $\textbf{C}\!=\!(0,0,n)$, as shown in Fig. \ref{f1}(b), each layer contributes a Chern number $\mathcal{C}_{xy}\!=\!n$, resulting in $nL_z$ chiral modes with quantized Hall conductances $G_{xy}\!=\!nL_ze^2/h$.
This framework has driven research on tunable quantum anomalous Hall effects, achieving significant experimental successes \cite{QAH11,QAH12,Mn5,3DQHCDW}. {However, Chern vectors with multiple nonzero components, which gives rise to more fascinating topological surface states and transport properties, remain largely unexplored.}

 Very recently, Chern vectors $\textbf{C}\!=\!(0,m,n)$ and their exotic surface states have been experimentally confirmed \cite{Chernvector,Chernvector2}.
  As shown in Fig. \ref{f1}(c), this configuration generates distinct chiral networks on the front and back surfaces, characterized by the Chern vectors and distinctly different from 2D Chern insulators \cite{Chernvector,Chernvector2}.
 While the chiral modes are still protected by the Chern numbers, the loops and links in the networks seem enable scattering between opposite surfaces, which deviates from the backscattering-free transport observed in $\textbf{C}\!=\!(0,0,n)$ systems \cite{QAH2}.
Consequently, conventional frameworks such as the two-terminal response formula $G_\alpha\!=\!\mathcal{C}_{\alpha\beta}L_\gamma e^2/h$ \cite{buttiker} and TKNN formula $G_{\alpha\beta}\!=\!\mathcal{C}_{\alpha\beta}L_\gamma e^2/h$ \cite{QH3}, may not be applicable.
This raises a key question: Do systems with Chern vector $\textbf{C}\!=\!(0,m,n)$ still exhibit any quantized transport response? If so, how are these responses related to the Chern vector? Addressing these questions is essential for understanding the topological significance of Chern vectors, particularly in their ability to predict and control transport phenomena in 3D topological systems.

Here, we unveil that a 3D quantized Hall effect is protected by the Chern vector $\textbf{C}=(0,m,n)$, bridging the gap between Chern vectors and their quantized transport properties.
The chiral networks induced by $\textbf{C}=(0,m,n)$ modify the transport characteristics of  chiral surface states, which deviate from the conventional two-terminal response formula $G_\alpha\!=\!\mathcal{C}_{\alpha\beta}L_\gamma$. (In this letter, all conductances are measured in units of $e^2/h$.)
We find that the quantized two-terminal response is reformulated as the sum of transmission coefficients along the $x$ and $y$-direction, given by $(mL_y\!+\!nL_z)$.
{This quantization underpins unique terminal-to-terminal responses, which can be directly measured from Hall bar setups. By solving the multi-terminal Landauer-B{\"u}ttiker formula \cite{LB1}, we predict four quantized Hall conductances} $G_{xy}\!=\!G_{xz}\!=\!-G_{yx}\!=\!-G_{zx}\!=\!(mL_y\!+\!nL_z)$. These conductances depend distinctly on sample dimensions, highlighting the 3D nature of the Hall effect. Finally, we propose several candidates, where these responses can be experimentally verified, providing a pathway to validate our theoretical predictions.

\B{\textit{Model and properties.---}}
Our starting point is the tight-binding Hamiltonian
\begin{equation}\label{Eq1}
H=\left(
\begin{array}{cc}
\mathcal{H}_{xyz} & 0\\
0& \mathcal{H}_{xzy}\\
\end{array}\right)+\sum_n\left(
\begin{array}{cc}
0 & T_c\\
T_c& 0\\
\end{array}\right)c_n^\dagger c_n,
\end{equation}
where $\mathcal{H}_{xyz}\!=\!\sum_n (T_0\!+\!\varepsilon_n) c^\dagger_nc_n\!+\!\sum_{\alpha\in x,y,z} T_\alpha c^\dagger_{n+\delta_\alpha}c_n\!+\!h.c.$ describes the 3D Chern insulator with layer Chern vector \(\textbf{C}=(\mathcal{C}_{yz}=0,\mathcal{C}_{xz}=0,\mathcal{C}_{xy}=1)\) \cite{QAH1,QAH2,QAH3,QAH6,QAH7,QAH8,QAH9,QAH10,QAH11,QAH12}. The hopping matrices are $T_0\!=\!\tau_z$, $T_x\!=\!M\tau_z/2-iA\tau_x/2$, $T_y\!=\!M\tau_z/2-iA\tau_y/2$, and $T_z\!=\!t_{1}\tau_z$. $c_n^\dagger$ ($c_n$) is the creation (annihilation) operator of site $n$.
The lattice constant in the $\alpha$-direction is represented by $\delta_\alpha$. A similar component, $\mathcal{H}_{xzy}$, is constructed with modified hopping terms: $T_y\!=\!t_{2}\tau_z$ and $T_z\!=\!M\tau_z/2-iA\tau_y/2$, characterized by $\textbf{C}=(\mathcal{C}_{yz}=0,\mathcal{C}_{xz}=1,\mathcal{C}_{xy}=0)$.

\begin{figure}[t]
\centering
\includegraphics[width=0.49\textwidth]{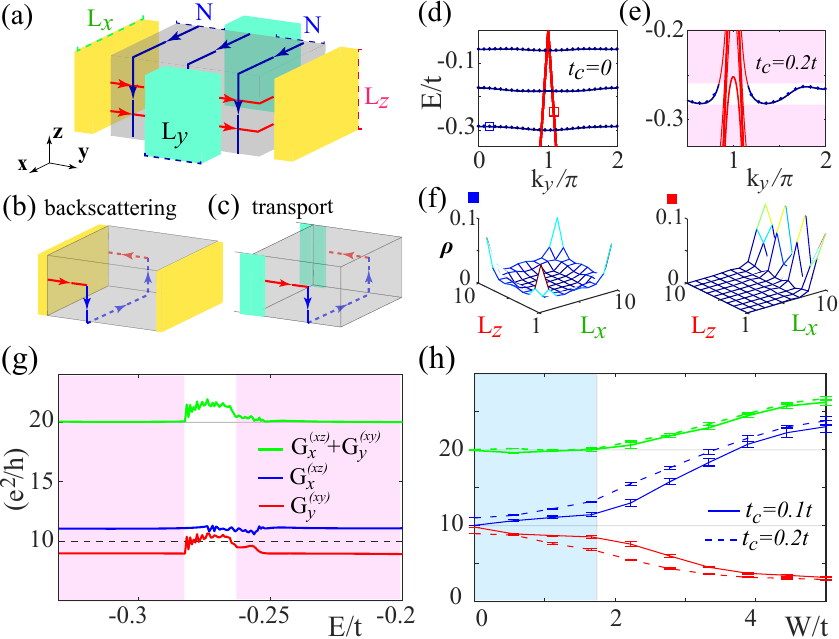}
	\caption{(a) Schematic distribution of the chiral surface states with Chern vector $\textbf{C}=(0,1,1)$. The sample sizes has been marked in the figure. (b) and (c) present the influences of chiral networks for different leads. (d) and (e) Spectra $E$ versus $k_y$ with and without $t_c$, respectively. (f) Typical plots of eigenstates $\rho=\!|\psi_n|^2$ for the blue and red states marked in (d). (g) $G^{(\alpha\beta)}_\gamma$ versus Fermi energy $E$. $G^{(\alpha\beta)}_\gamma$ denotes the two-terminal conductance contributed by chiral modes located on the $\alpha$-$\beta$ boundaries with current along the $\gamma$ direction. Here, $\alpha,\beta,\gamma\!\in\!(x,y,z)$. (h) Surface conductance $G^{(\alpha\beta)}_\gamma$ versus disorder strength $W$ with Fermi energy $E\!=\!-0.32t$. The other parameters are $L_x\!=\!L_y\!=\!L_z\!=\!10$ and $N\!=\!15$. }
	\label{f2}
\end{figure}

According to the bulk-boundary correspondence, red chiral modes from  $\mathcal{H}_{xyz}$ emerge along $x$-$y$ boundaries, while blue modes from $\mathcal{H}_{xzy}$ appear along $x$-$z$ boundaries, as illustrated in Fig. \ref{f2}(a). Their combination, denoted as $\mathcal{H}$, naturally exhibits a composite Chern vector $\textbf{C}=(\mathcal{C}_{yz}=0,\mathcal{C}_{xz}=1,\mathcal{C}_{xy}=1)$, with the copuling facilitated by the hopping matrix $T_c\!=\!t_c(\textbf{I}_{2\times 2}+\tau_x)$ [see SM.I for details \cite{SM}].

For subsequent calculations, the parameters are fixed as  $A\!=\!t$, $M\!=\!2t$, $t_{1}\!=\!-0.3t$ and $t_{2}\!=\!-0.1t$. Anderson disorder is also incorporated as onsite potential fluctuations, where $\varepsilon_n$ is uniformly distributed within $[-W/2,W/2]$, with $W$ representing the disorder strength \cite{Anderson0,Anderson1,Anderson2}.

\B{\textit{Two-terminal responses.---}} To clearly capture the topological features, we examine the influence of $t_c$ on the two-terminal responses. Here, we introduce the notation $G^{(\alpha\beta)}_\gamma$ to denote the two terminal conductance in the $\gamma$-direction, originating from the chiral modes located on the $\alpha$-$\beta$ boundaries. For example, in Fig. \ref{f1}(b), only the red modes on the $x$-$y$ boundaries contribute to the $x$-direction conductance, which can thus expressed as $G^{(xy)}_x=\mathcal{C}_{xy}L_z=nL_z$.

In the decoupled limit ($t_c\!=\!0$) of $H$, the red and blue chiral modes independently contribute to the two-terminal conductance. In the $y$ direction, the red modes exhibit traveling feature, while blue modes are nearly localized in the $y$-direction, forming relatively flat bands [see Fig s. \ref{f2}(d)]. Consequently, the $y$-direction conductance arises solely from the red modes, given by $G^{(xy)}_y=\mathcal{C}_{xy}L_z$ [conductance between yellow leads in Fig. \ref{f2}(a)]. Similarly, in the $x$ direction, the conductance contributed by the blue modes on the $x$-$z$ boundaries satisfy $G^{(xz)}_x=\mathcal{C}_{xz}L_y$.

However, as the coupling is turned on ($t_c\neq0$), the two-terminal responses are not topologically protected any more, in contrasted to the case of 2D Chern insulators. Although the components of the Chern vector remain invariant as per the TKNN formula [see SM.I], the coupling alters the transport behavior. In the $y$ direction, the red modes coupled with the blue ones, partially gapping out some of them [see Fig. \ref{f2}(e)]. As a result, $G^{(xy)}_y$ deviates from $\mathcal{C}_{xy}L_z$. For instance, in both the clean case [Figs. \ref{f2}(g)] and the disordered case [Figs. \ref{f2}(h)], $G^{(xy)}_y$ is smaller than the value $\mathcal{C}_{xy}L_z=10$. Intuitively, this reduction in $G^{(xy)}_y$ arises from backscattering induced by the coupling between chiral modes on opposite surfaces, as shown in Fig. \ref{f2}(b).

In the $x$-direction, $G^{(xz)}_x$ also deviates from $\mathcal{C}_{xz}L_y$, but to a higher value, in both the clean case [Figs. \ref{f2}(g)] and the disordered case [Figs. \ref{f2}(h)]. This increase stems from additional transport channels enabled by the coupling [see Fig. \ref{f2}(c)]. Thus, while the two-terminal responses $G^{(xy)}_y$ and $G^{(xz)}_x$ are no longer protected by the topology, the Chern vector $(0,1,1)$ indicates the existence of robust transport quantities that warrant further exploration.

Remarkably, the processes that decrease $G^{(xy)}_y$ [Fig. \ref{f2}(b)] and increase $G^{(xz)}_x$ [Fig. \ref{f2}(c)] are fundamentally identical. This observation suggests that the sum $G^{(xz)}_x\!+\!G^{(xy)}_y$ may remain topologically protected [see SM.IV for more details \cite{SM}]. Specifically,
\begin{equation}
G^{(xz)}_x\!+\!G^{(xy)}_y\!=\!\mathcal{C}_{xz}L_y+\mathcal{C}_{xy}L_z.
\end{equation}
To verify this conclusion, Fig. \ref{f2}(g) shows numerical results demonstrating that $G^{(xz)}_x\!+\!G^{(xy)}_y$ is quantized to $20$ ($20=\mathcal{C}_{xz}L_y+\mathcal{C}_{xy}L_z$) within the pink energy interval \cite{pinkenergyinterval}. Furthermore, the robustness of this topological quantity is confirmed in Fig. \ref{f2}(h), where $G^{(xz)}_x\!+\!G^{(xy)}_y$ remains quantized for significant disorder strengths and varying $t_c$, even though both $G^{(xz)}_x$ and $G^{(xy)}_y$ are fragile.

Similarly, the $x$-direction conductance $G_x=G^{(xz)}_x\!+\!G^{(xy)}_x$ emerges as an extra topologically protected response [see SM.III for details \cite{SM}].

\B{\textit{Multi-terminal Hall responses.---}}
To demonstrate that the quantized conductance $(\mathcal{C}_{xz}L_y+\mathcal{C}_{xy}L_z)$ can be directly measured using Hall bar setups, we investigate the quantum transport of a sample with $\textbf{C}=(\mathcal{C}_{yz}=0,\mathcal{C}_{xz}=1,\mathcal{C}_{xy}=1)$ through the Landauer-B{\"u}ttiker formalism \cite{LB1,SM}. The general relationship between current and voltage can be expressed as \cite{LB1,SM}:
\begin{equation}\label{LBequation}
I_i=\frac{e^2}{h}\sum_{j\neq i}T_{ij}(V_i-V_j),
\end{equation}
where $V_i$ represents the voltage at the $i$-th lead, $I_i$ denotes the current flowing out of the $i$-th lead into the sample, and $T_{ij}$ is the transmission probability from the $j$-th to the $i$-the lead.

\begin{figure}[t]
\centering
\includegraphics[width=0.49\textwidth]{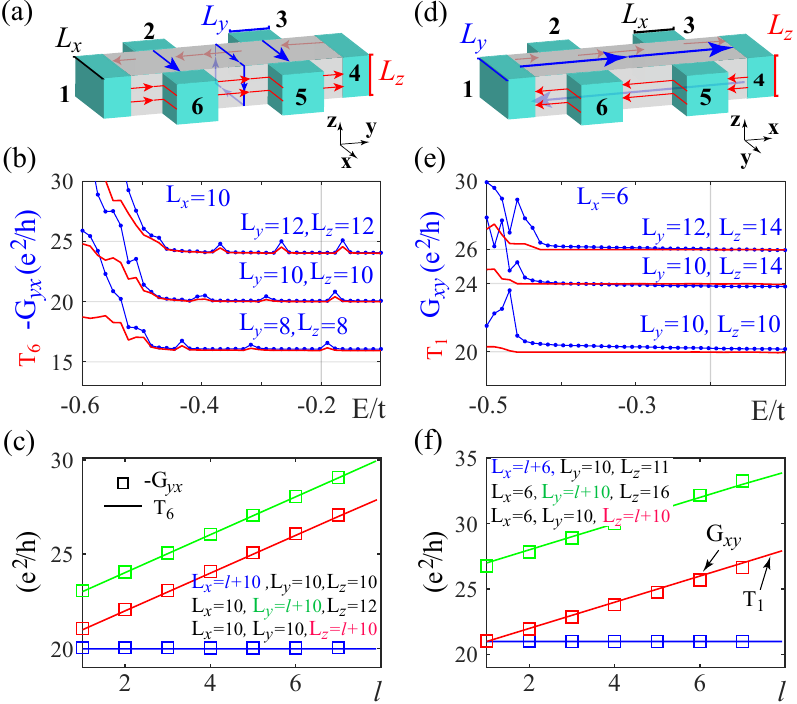}
	\caption{(a) The schematic plot of Hall bar with current along the $y$-direction. The sample sizes have been marked in the figure with $t_c=0.1t$. (b) and (c) the Hall conductances versus Fermi energy $E$ and sample size for the setup in (a). (d)-(f) are the same as those in (a)-(c), excepting currents along the $x$-direction. We set $E\!=\!-0.2t$ for (c) and (f).}
	\label{f3}
\end{figure}

\textbf{Hall bar along the $y$-direction:} For a standard six-terminal Hall bar as shown in Fig. \ref{f3}(a), current flows between leads 1 and 4, while leads 2, 3, 5 and 6 measure the voltages. The red and blue chiral modes yield the following nonzero transmission matrix elements:
\begin{equation}
\begin{array}{l}
T_{j,j+1}=T_{61}=G^{(xy)}_y,\\
T_{26/35}=T_{62/53}=G^{(xz)}_x,
\end{array}
\end{equation}
where $j\in[1,2,3,4,5]$. Defining $T_6=\sum_{j\neq6}T_{6,j}$, one obtains $T_6=T_{61}+T_{62}=G^{(xy)}_y+G^{(xz)}_x$, which is quantized as $(\mathcal{C}_{xz}L_y+\mathcal{C}_{xy}L_z)$. Solving the Eq. (\ref{LBequation}) yields the quantized Hall conductance between leads 2 and 6 [see SM.V for more details \cite{SM}]
\begin{equation}\label{Hallyx}
\begin{array}{l}
G_{yx}\!=\!I_y/(V_2\!-\!V_6)\!=\!-T_6\!=\!-(\mathcal{C}_{xz}L_y\!+\!\mathcal{C}_{xy}L_z).
\end{array}
\end{equation}
where $I_y$ is the current along the $y$-direction.

\textbf{Hall bar along the $x$-direction:} Due to the 3D nature of the Chern vector, the Hall bar can also be fabricated along the $x$-direction, as shown in Fig. \ref{f3}(d). The chirality of red modes, associated with the nonzero Chern component $\mathcal{C}_{xy}$, results in the transmission matrix elements $T_{j+1,j}=T_{1,6}=G^{(xy)}_x$ for $j\in[1,2,3,4,5]$. Meanwhile, the blue modes from the nonzero Chern component $\mathcal{C}_{xz}$ connect leads 1 and 4 with $T_{14}=T_{41}=G^{(xz)}_x$. Defining $T_1=\sum_{j\neq1}T_{1,j}$, we find $T_1=T_{16}+T_{14}=G^{(xy)}_x+G^{(xz)}_x$, which is quantized as $(\mathcal{C}_{xz}L_y+\mathcal{C}_{xy}L_z)$. The Hall conductance measured between leads 2 and 6 is then quantized as
\begin{equation}\label{Hallxy}
\begin{array}{l}
G_{xy}\!=\!I_x/(V_2-V_6)\!=\!+T_1\!=\!(\mathcal{C}_{xz}L_y+\mathcal{C}_{xy}L_z).
\end{array}
\end{equation}

The validity of Eq. (\ref{Hallyx}) and Eq. (\ref{Hallxy}) can be verified by numerical calculations using the non-equilibrium Green's function technique \cite{LB1}. As shown in Fig. \ref{f3}(b), the Hall conductance $G_{yx}$ aligns with the quantized value $T_1=(\mathcal{C}_{xz}L_y+\mathcal{C}_{xy}L_z)$ across varying sample sizes. A similar agreement is observed for $G_{xy}$ in Fig. \ref{f3}(e).

Unlike the single sample-size dependence of $G_{xy}\!=\!nL_z$, typically observed for $\textbf{C}\!=\!(0,0,n)$ [Fig. \ref{f1}(b)], the 3D quantized Hall conductance of $\textbf{C}\!=\!(0,m,n)$ exhibits double sample-size dependence and can be observed on different directions. Specifically, $G_{yx}$ and $G_{xy}$ depend on both $L_x$ and $L_y$, as clearly shown in Fig. \ref{f3}(c) and (f). This double-size dependence serves as a distinctive experimental fingerprint of the proposed 3D quantized Hall effects protected by the Chern vector.

\B{\textit{Candidate I: photonic and acoustic crystals.---}}
Below, we discuss several experimental candidates exhibiting a Chern vector \( \mathbf{C} = (0, m, n) \). Notably, \( \mathbf{C} = (0, m, n) \) has been successfully realized in both photonic and acoustic crystals \cite{Chernvector, Chernvector2}. Although it is difficult to capture the Hall responses in these systems, the characteristic backscattering trajectories of the chiral surface states, as shown in Figs.~\ref{f2}(b) and \ref{f2}(c), can be experimentally validated by tracking the propagation of wavepackets. These distinct trajectories form the basis of the quantized two-terminal response
$
G^{(xz)}_x + G^{(xy)}_x = (mL_y + nL_z),
$
which underpins the quantized Hall conductances described in Eqs.~(\ref{Hallyx}) and (\ref{Hallxy}).

\B{\textit{Candidate II: Chern vector protected 3D QHE.---}}
We next propose candidates for realizing the Chern vector \( \mathbf{C} = (0, m, n) \) in condensed matter systems with measurable Hall responses. Three-dimensional (3D) Chern insulator bands have been identified in various materials \cite{QAH11, QAH12, Mn5, 3DQHCDW}. Considering the common coexistence of metallic bands in topological systems, we analyze a 3D Chern insulator coupled to metallic bands, described by the Hamiltonian:
\begin{equation}\label{Eq2}
	H\!=\!\left(
	\begin{array}{cc}
	\mathcal{H}_\text{metal}(\textbf{B}) & 0\\
	0&  \mathcal{H}_{xzy}(\textbf{B})\\
	\end{array}\right)\!+\!\sum_n\left(
	\begin{array}{cc}
	0 & t_c\\
	t_c& 0\\
	\end{array}\right)c_n^\dagger c_n.
\end{equation}
Here, the metallic bands are described by:
$
\mathcal{H}_{\text{metal}}(\mathbf{B}) = \sum_n \varepsilon_0 c^\dagger_n c_n
+ t_x e^{i\phi y} c_{n+\delta_x}^\dagger c_n
+ t_y c_{n+\delta_y}^\dagger c_n
+ t_z c_{n+\delta_z}^\dagger c_n + \text{h.c.},
$
and $ \mathcal{H}_{xzy}(\mathbf{B}) $ is identical to Eq.~(\ref{Eq1}), with the inclusion of an applied magnetic field $ \mathbf{B} $.
The magnetic field \( \mathbf{B} = (0, 0, B) \) is incorporated by introducing an extra flux \( \phi = a^2B/\phi_0 \) through the unit lattice area \( \delta_x \times \delta_y = a^2 \) \cite{LB1}. In our calculations, the parameters are set as \( t_x = t_y = t \), \( \varepsilon_0 = -2t \), and \( t_z = 0.1t \).

\begin{figure}[t]
	\centering
	\includegraphics[width=0.48\textwidth]{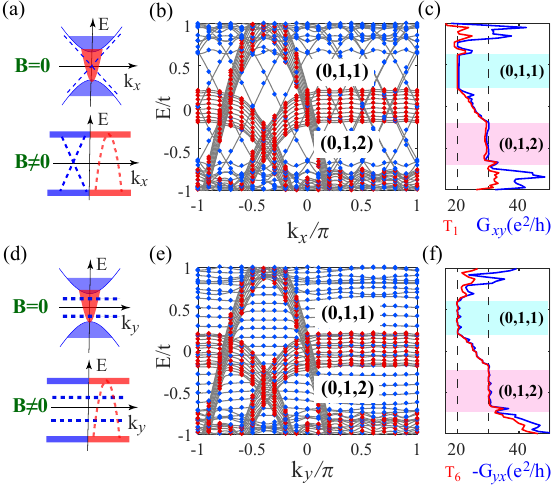}
	\caption{(a) and (d): Schematic band spectra along $k_x$ and $k_y$. The blue (red) solid bands mark the bulk bands of 3D Chern insulator (metallic bands). The dashed lines are the corresponding chiral surface states. (b) and (e): the numerically calculated band spectra with $\phi\!=\!0.25\pi$. The blue (red) circles roughly mark the 3D Chern insulator (3D QHE) bands in (a) and (d). (c) and (f) are the Hall conductances of (b) and (e), respectively. We fix $L_x\!=\!L_y\!=\!L_z\!=\!10$ and $t_c\!=\!0.1t$.
	 }
	\label{f4}
\end{figure}

As schematically illustrated in Figs.~\ref{f4}(a) and \ref{f4}(d), the 3D Chern insulator coupled with metallic bands can achieve the Chern vector \( \mathbf{C} = (0, m, n) \). For \( \mathbf{B} \neq 0 \), the chiral surface states of the 3D Chern insulator retain an open Fermi surface, which prevents Landau quantization of the chiral states \cite{3DQHE1, 3DQHE2}. The magnetic field simply shifts their momentum without affecting their chirality, as depicted in Fig.~\ref{f4}(a). Consequently, \( \mathcal{H}_{xzy}(\mathbf{B}) \) still produces a valid, \( \mathbf{B} \)-independent Chern vector component \( \mathbf{C} = (0, 1, 0) \).

More importantly, the metallic bands, represented by the red states in Figs.~\ref{f4}(a) and \ref{f4}(d) and described by \( \mathcal{H}_{\text{metal}}(\mathbf{B}) \), contribute to the 3D quantum Hall effect (QHE) when \( \mathbf{B} \neq 0 \) \cite{3DQHCDW}. By applying a magnetic field \( \mathbf{B} \) that is not parallel to the Chern vector \( \mathbf{C} = (0, m, 0) \), it becomes possible to achieve the desired Chern vector \( \mathbf{C} = (0, m, n) \).

For clarity, we numerically calculated the spectra of Eq.~(\ref{Eq2}) with \( \mathbf{B} \neq 0 \). The numerically obtained surface bands of the 3D Chern insulator exhibit splitting along \( k_x \), as shown in Fig.~\ref{f4}(b), consistent with the analysis illustrated in Fig.~\ref{f4}(a).
Similarly, the numerical results for \( \mathcal{H}_{\text{metal}}(\mathbf{B}) \), presented in Figs.~\ref{f4}(b) and \ref{f4}(e), reveal the expected features, confirming the existence of the 3D quantum Hall effect (QHE) with \( \mathbf{C} = (0, 0, n) \). Consequently, \( H \) in Eq.~(\ref{Eq2}) exhibits the Chern vectors \( \mathbf{C} = (0, 1, 1) \) and \( \mathbf{C} = (0, 1, 2) \) at different Fermi energies.

To further verify the Hall responses, we considered samples with dimensions \( L_x = L_y = L_z = 10 \). As shown in Figs.~\ref{f4}(c) and \ref{f4}(f), the calculated Hall conductances align with the predicted formula in Eq.~(\ref{Hallyx}) and Eq.~(\ref{Hallxy}). Specifically, for \( \mathbf{C} = (0, 1, 2) \), the Hall conductance is \( G_{xy} = -G_{yx} = 30 \), and for \( \mathbf{C} = (0, 1, 1) \), it is \( G_{xy} = -G_{yx} = 20 \).

Notably, \( T_1 \) (\( T_6 \)) overlaps with \( G_{xy} \) (\(-G_{yx}\)) for Fermi energies \( E \in [-0.1t, 0.1t] \), which lie within the bulk bands of \( \mathcal{H}_{\text{metal}}(\mathbf{B}) \). This observation indicates that the proposed Hall effect scheme remains valid in these energy regions, where the bulk band gaps are replaced by finite-size energy gaps. Such a replacement modifies the integer Chern vector coefficient \( m \) to fractional values, \( m \in [1, 2] \). This relaxation of the Chern vector requirement facilitates the realization of \( \mathbf{C} = (0, m, n) \) and the corresponding quantized Hall response \( \pm(mL_y + nL_z) \) in experiments.

\B{\textit{Summary and Discussion.---}}
In summary, we have extended the concept of Chern-number-protected quantized Hall responses in 3D systems to cases characterized by the Chern vector \( \mathbf{C} = (0, m, n) \). We have demonstrated that the sum of transmission coefficients along both the \( x \)- and \( y \)-directions is protected by the Chern vector, preserving the quantized value \( (mL_y + nL_z) \). This summation relation leads to the quantized Hall conductances \( \pm(mL_y + nL_z) \).

A distinctive feature of these Hall conductances is their peculiar double sample-size dependence, which serves as a unique experimental fingerprint. To validate our theoretical predictions, we propose two typical candidates for experimental realization:
(1) Photonic or acoustic 3D Chern insulators with \( \mathbf{C} = (0, m, n) \);
(2) 3D Chern insulators with metallic bands and magnetic fields in condensed matter systems.

Our study advances the understanding of the Hall effect in 3D systems. Notably, the quantized Hall conductance \( G_{xy} = \kappa L_z  \) observed in 3D systems may not directly correspond to \( \mathbf{C} = (0, 0, \kappa) \). Instead, it can originate from a Chern vector \( \mathbf{C} = (0, m, n) \), where \( \kappa = (mL_y + nL_z)/L_z \). Furthermore, exploring transport properties for arbitrary Chern vectors \( \mathbf{C} = (\mathcal{C}_{yz}, \mathcal{C}_{xz}, \mathcal{C}_{xy}) \) presents an exciting avenue for future investigations, potentially broadening the knowledge of the Hall effect in 3D systems.

\textit{ Acknowledgements.---}
We appreciate the valuable discussions with Hong Yao, Li-Yuan Zhang, Ming Lu, Ming Gong and Qiang Wei. This work was supported by the National Basic Research Program of China (Grants No. 2024YFA1409003 and No. 2022YFA1403700), National Natural Science Foundation of China (Grants No. 12350401 and No. 12204044).


\begin{thebibliography}{99}
\bibitem{zongshu1}M. Z. Hasan and C. L. Kane, Colloquium: topological insulators, Rev. Mode. Phys. \textbf{82}, 3045 (2010).
\bibitem{zongshu2}X. L. Qi and S. C. Zhang, Topological insulators and superconductors, Rev. Mod. Phys. \textbf{83}, 1057 (2011).
\bibitem{shen1}S. Q. Shen, {\it Topological insulators}, Berlin, Springer (2012).

\bibitem{QH1}V. K. Klitzing, G. Dorda and M. Pepper, New Method for High-Accuracy Determination of the Fine-Structure Constant Based on Quantized Hall Resistance, Phys. Rev. Lett. \textbf{45}, 494 (1980).
\bibitem{QH2}R. B. Laughlin, Quantized Hall conductivity in two dimensions, Phys. Rev. B \textbf{23}, 5632 (1981).
\bibitem{QH3}D. J. Thouless, M. Kohmoto, M. P. Nightingale, and M. den Nijs, Quantized Hall Conductance in a Two-Dimensional Periodic Potential, Phys. Rev. Lett. \textbf{49}, 405 (1982).


\bibitem{QAH1}B. I. Halperin, Quantized Hall conductance, current-carrying edge states, and the existence of extended states in a two-dimensional disordered potential, Phys. Rev. B \textbf{25}, 2185 (1982).
\bibitem{QAH2}C. Z. Chang, et al., Experimental observation of the quantum anomalous Hall effect in a magnetic topological insulator, Science \textbf{340}, 167 (2013).
\bibitem{QAH3}F. D. M. Haldane, Model for a quantum Hall effect without Landau levels: Condensed-matter realization of the parity anomaly, Phys. Rev. Lett. \textbf{61}, 2015 (1988).
\bibitem{QAH6}E. Prodan, Disordered topological insulators: a non-commutative geometry perspective, Phys. A-Math. Theor. \textbf{44}, 113001 (2011).
\bibitem{QAH7}Y. Gong, et al., Experimental realization of an intrinsic magnetic topological insulator, Chin. Phys. Lett. \textbf{36}, 076801 (2019).

\bibitem{QAH8} H. Park, et al.,Observation of fractionally quantized anomalous Hall effect, Nature \textbf{622},74 (2023).
\bibitem{QAH9} Z. G. Lu, et al., Fractional quantum anomalous Hall effect in multilayer graphene, Nature \textbf{626},759 (2024).

\bibitem{QAH10}  F. Xu, Z. Sun, T. Jia, C. Liu, C. Xu, C. S. Li, et al., Observation of Integer and Fractional Quantum Anomalous Hall Effects in Twisted Bilayer MoTe$_2$, Phys. Rev. X \textbf{13}, 031037 (2023).

\bibitem{QAH11} Y. F. Zhao, et al.,Tuning the Chern number in quantum anomalous Hall insulators, Nature \textbf{588}, 419 (2020).
\bibitem{QAH12} Y. F. Zhao, R. X. Zhang, L. J. Zhou, R. B. Mei, Z. J. Yan, M. H. W. Chan, C. X. Liu, and C. Z. Chang, Zero Magnetic Field Plateau Phase Transition in Higher Chern Number Quantum Anomalous Hall Insulators, Phys. Rev. Lett. \textbf{128}, 216801 (2022).
\bibitem{Mn5}M. M. Otrokov, et al., Unique Thickness-Dependent Properties of the vander Waals Interlayer Antiferromagnet MnBi$_2$Te$_4$ Films, Phys. Rev. Lett. \textbf{122}, 107202 (2019).
\bibitem{3DQHCDW}  F. Tang, et al. Three-dimensional quantum Hall effect and metal-insulator transition in ZrTe$_5$, Nature \textbf{569}, 537 (2019).
\bibitem{3DQHE1}C. M. Wang, H. P. Sun, H. Z. Lu, and X. C. Xie, 3D quantum Hall effect of Fermi arcs in topological semimetals, Phys. Rev. Lett. \textbf{119}, 136806 (2017).
\bibitem{3DQHE2}H. L. Li, H. W. Liu, H. Jiang,  and X. C. Xie, 3D quantum Hall effect and a global picture of edge states in Weyl semimetals, Phys. Rev. Lett. \textbf{125}, 036602 (2020).
 \bibitem{3DQHE3}C. Zhang, et.al., Quantum Hall effect based on Weyl orbits in Cd3As2, Nature \textbf{565}, 331-336 (2019).
 \bibitem{3DQHE4}C. Zhang, et. al., Evolution of Weyl orbit and quantum Hall effect in Dirac semimetal Cd3As2, Nat. Commun. \textbf{8}, 1-8 (2017).
 \bibitem{3DQHE5}T. Schumann, L. Galletti, D. Kealhofer, H. Kim, M. Goyal, and S. Stemmer, Observation of the quantum Hall effect in confined films of the three-dimensional Dirac semimetal Cd3As2, Phys. Rev. Lett. \textbf{120}, 016801 (2018).

\bibitem{3DHE1} L. Li, J. Cao, C. Cui, Z.-M. Yu and Y. Yao, Planar Hall effect in topological Weyl and nodal-line semimetals, Phys. Rev. B \textbf{108}, 085120 (2023).
\bibitem{3DHE2} Z. Z. Du, H.-Z. Lu and X. C. Xie, Nonlinear Hall effects, Nat. Rev. Phys. \textbf{3}, 744 (2021).
\bibitem{3DHE3} J. X. Yin et al., Quantum-limit Chern topological magnetism in TbMn$_6$Sn$_6$. Nature \textbf{583}, 533 (2020).
\bibitem{3DHE4} N. Shumiya et al., Evidence of a room-temperature quantum spin Hall edge state in a higher-order topological insulator, Nature Materials \textbf{21}, 1111 (2022).
\bibitem{3DHE5}M. S. Hossain, et al., Quantum transport response of topological hinge modes, Nat. Phys. \textbf{20}, 776 (2024).



\bibitem{Chernvector}G. G. Liu et. al., Topological Chern vectors in three-dimensional photonic crystals, Nature \textbf{609}, 925 (2022).
\bibitem{Chernvector2} L. Yang et. al., Acoustic Three-dimensional Chern Insulators with Arbitrary Chern Vectors, arXiv:2401.07040v1 (2024).
\bibitem{Chernvector3}F. Ma, J. Feng, F. Li, Y. Wu, and D. Zhou, Floquet Chern Vector Topological Insulators in Three Dimensions, arXiv:2412.00619v1 (2024).
\bibitem{buttiker}M. B{\"u}ttiker, Edge-State Physics Without Magnetic Fields, Science, 325, 5938 (2009);
\bibitem{LB1}S. Datta, {\it Electronic Transport in Mesoscopic Systems}, Cambridge University Press, Cambridge, England, (1995)


\bibitem{SM} See Supplementary Materials for more details.



\bibitem{Anderson0}P. W. Anderson, Absence of diffusion in certain random lattices, Phys. Rev. \textbf{109}, 1492 (1958).
\bibitem{Anderson1}E. Abrahams, P. W. Anderson, D. C. Licciardello, and T. V. Ramakrishnan, Scaling theory of localization: Absence of quantum diffusion in two dimensions, Phys. Rev. Lett. \textbf{42}, 673 (1979).
\bibitem{Anderson2}F. Evers, and A. D. Mirlin, Anderson transitions, Rev. Mod. Phys. \textbf{80}, 1355 (2008).
\bibitem{pinkenergyinterval} Actually, the white energy interval possess the trivial surface states along the $y$ direction. These states originate from the stacking of chiral modes along the $y$ direction. Noevertheless, the white energy interval is rather tiny, which can be roughly neglected.
\end{thebibliography}
\end{document}


\title{Supplementary Materials for ``Chern Vector Protected Three-dimensional Quantized Hall effects"}

\author{Zhi-Qiang Zhang }
\affiliation{Interdisciplinary Center for Theoretical Physics and Information Sciences (ICTPIS), Fudan University, Shanghai 200433, China}
\affiliation{School of Physical Science and Technology, Soochow University, Suzhou 215006, China}
\author{Shu-Guang Cheng }
\affiliation{Department of Physics, Northwest University, Xi'an 710069, China}
\author{Hongfang Liu }
\affiliation{School of Physical Science and Technology, Soochow University, Suzhou 215006, China}
\author{Hailong Li }\email{hailongli@fudan.edu.cn}
\affiliation{Interdisciplinary Center for Theoretical Physics and Information Sciences (ICTPIS), Fudan University, Shanghai 200433, China}
\author{Hua Jiang}\email{jianghuaphy@fudan.edu.cn}
\affiliation{Interdisciplinary Center for Theoretical Physics and Information Sciences (ICTPIS), Fudan University, Shanghai 200433, China}
\author{X. C. Xie}
\affiliation{Interdisciplinary Center for Theoretical Physics and Information Sciences (ICTPIS), Fudan University, Shanghai 200433, China}
\affiliation{International Center for Quantum Materials, School of Physics, Peking University, Beijing 100871, China}
\affiliation{Hefei National Laboratory, Hefei 230088, China}

\maketitle
\tableofcontents


\section{I. Chern vector}

In this section, we present the Chern vector $\textbf{C}=(\mathcal{C}_{yz},\mathcal{C}_{xz},\mathcal{C}_{xy})$ more clearly.
Based on the tight-bounding model in the main text, one has
\begin{equation}\label{Eqss1}
\mathcal{H}\!=\!\left(
\begin{array}{cc}
\mathcal{H}_{xyz} & 0\\
0& \mathcal{H}_{xzy}\\
\end{array}\right)+\sum_n\left(
\begin{array}{cc}
0 & T_c\\
T_c& 0\\
\end{array}\right)c_n^\dagger c_n.
\end{equation}
$\mathcal{H}_{xyz}=\sum_n (T_0+\varepsilon_n) c^\dagger_nc_n+\sum_{\alpha\in[x,y,z]} T_\alpha c^\dagger_{n+\delta_\alpha}c_n+h.c.$ describes the 3D Chern insulators with layer Chern number $\mathcal{C}_{xy}$ in the $x$-$y$ plane. The hoppings satisfy $T_0=\tau_z$, $T_x=M\tau_z/2-iA\tau_x/2$, $T_y=M\tau_z/2-iA\tau_y/2$, and $T_z=t_{z1}\tau_z$. Here, $T_c=t_c(\textbf{I}+\tau_x)$ with $
[\textbf{I}+\tau_x]=
\left(
\begin{array}{cc}
1 & 1\\
1& 1\\
\end{array}
\right)
$. Such a formula fully considered the coupling between different orbital degeneracies.
After neglecting the disorder $\varepsilon_n$, $\mathcal{H}_{xyz}$ can be rewritten as:
\begin{align}
	\begin{split}
\mathcal{H}_{xyz}=\sum_n \tau_z c^\dagger_nc_n
+ (M\tau_z/2-iA\tau_x/2) c^\dagger_{n+\delta_x}c_n
+ (M\tau_z/2-iA\tau_y/2) c^\dagger_{n+\delta_y}c_n
+ (t_{z1}\tau_z) c^\dagger_{n+\delta_z}c_n
  +h.c.
	\end{split}
\end{align}
Its momentum formula reads:
\begin{align}
	\begin{split}
\mathcal{H}_{k_xk_yk_z}=\tau_z
+[M\tau_z\cos k_x+A\tau_x\sin k_x]
+[M\tau_z\cos k_y+A\tau_y\sin k_y]
+[2t_{z1}\tau_z\cos k_z]
	\end{split}
\end{align}
Similar, one has:
 \begin{align}
	\begin{split}
\mathcal{H}_{k_xk_zk_y}=\tau_z
+[M\tau_z\cos k_x+A\tau_x\sin k_x]
+[2t_{z2}\tau_z\cos k_y]
+[M\tau_z\cos k_z+A\tau_y\sin k_z].
	\end{split}
\end{align}
Thus, one has:
\begin{equation}
\mathcal{H}(\textbf{K})=
\left(
\begin{array}{cc}
\mathcal{H}_{k_xk_yk_z} & 0\\
0 & \mathcal{H}_{k_xk_zk_y}\\
\end{array}
\right)+
\left(
\begin{array}{cc}
0 & t_c[\textbf{I}+\tau_x]\\
t_c[\textbf{I}+\tau_x] & 0\\
\end{array}
\right).
\end{equation}

 \begin{figure*}[th]
\includegraphics[width=16cm]{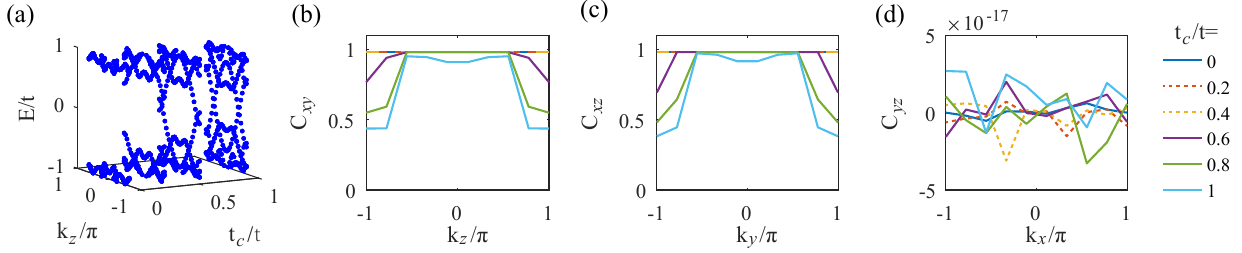}
\caption{(a) The bulk eigenvalues versus $k_z$ and $t_c$. (b)-(c) The Chern numbers $\mathcal{C}_{\alpha\beta}(k_\gamma)$ for different $k_\gamma$ and $t_c$. Here, $\alpha,\beta,\gamma\in x,y,z$. The other parameters are the same as those in the main text.
 }
\label{Sanalytical}
\end{figure*}

The bulk bands versus $k_z$ for different $t_c$ are are plotted in Figs. \ref{Sanalytical}(a), where The band gap holds while $t_c<0.5t$.
To clarify their Chern vectors, the standard Thouless-Kohmoto-Nightingale-Nijs (TKNN) formula \cite{chern3} are adopted, with
\begin{equation}
\mathcal{C}_{\alpha\beta}(k_\gamma)=\frac{i}{2\pi}\iint_{T^2}\sum_{E_c,E_v}\frac{\langle u_c|\partial_{k_\alpha} \mathcal{H}(\textbf{K})|u_v\rangle\langle u_v|\partial_{k_\beta} \mathcal{H}(\textbf{K}) k_\beta|u_c\rangle-\langle u_c|\partial_{k_\beta} \mathcal{H}(\textbf{K})|u_v\rangle\langle u_v|\partial_{k_\alpha} \mathcal{H}(\textbf{K})|u_c\rangle}{(E_c-E_v)^2}dk_\alpha dk_\beta.
\end{equation}
Here, $\alpha,\beta,\gamma\in x,y,z$.
$u_c$ and $u_v$ are the conduction and valence bands with eigenvalues $E_c<0$ and $E_v>0$, respectively.

As shown in Figs. \ref{Sanalytical}(b)-(d), the Chern numbers illuminate the Chern vector $\textbf{C}=(\mathcal{C}_{yz},\mathcal{C}_{xz},\mathcal{C}_{xy})=(0,1,1)$ for $k_{x/y/z}\in[-\pi,\pi]$. The Chern numbers $(\mathcal{C}_{yz},\mathcal{C}_{xz},\mathcal{C}_{xy})=(0,1,1)$ holds even $t_c\neq0$. Significantly, the typical quantized values for $k_{x/y/z}\in[-\pi,\pi]$ deviate from their nontrivial values when the bulk band gap closed, where $t_c\approx0.5t$.

\section{ II.  Landauer-B{\"u}ttiker formula}

In our calculation, the transmission coefficient $T_{nm}$ and two terminal conductance $G_\alpha=T_{nm}e^2/h$ is obtained by adopting the Landauer-B{\"u}ttiker formula \cite{LB1}:
\begin{equation}
T_{nm}={\rm Tr}[\Gamma_nG^r\Gamma_mG^a].
\end{equation}
The retarded Green's function satisfies $G^r=[E+i\eta-\mathcal{H}-\Sigma_n-\Sigma_m]^{-1}$ for the two-terminal systems.
$\mathcal{H}$ is presented in Eq. (\ref{Eqss1}). $\Gamma_{n/m}=i[\Sigma_{n/m}-\Sigma^\dagger_{n/m}]$ is the linewidth function of the leads and $\Sigma_{n/m}$
is the self-energy of the leads. In our calculations, we set $\Sigma_{n/m}=-i\mathcal{I}_{L_\alpha\times L\beta}$, representing the metallic lead. Here $\mathcal{I}_{L_\alpha\times L_\beta}$ is an unit matrix with size $L_\alpha\times L_\beta$. $L_\alpha$ and $L_\beta$ are the sample sizes of the leads along $\alpha$ and $\beta$ directions. The self-energy for other kinds of metallic leads should not show qualitative influences to the final results in general.

 \begin{figure*}[h]
\includegraphics[width=17cm]{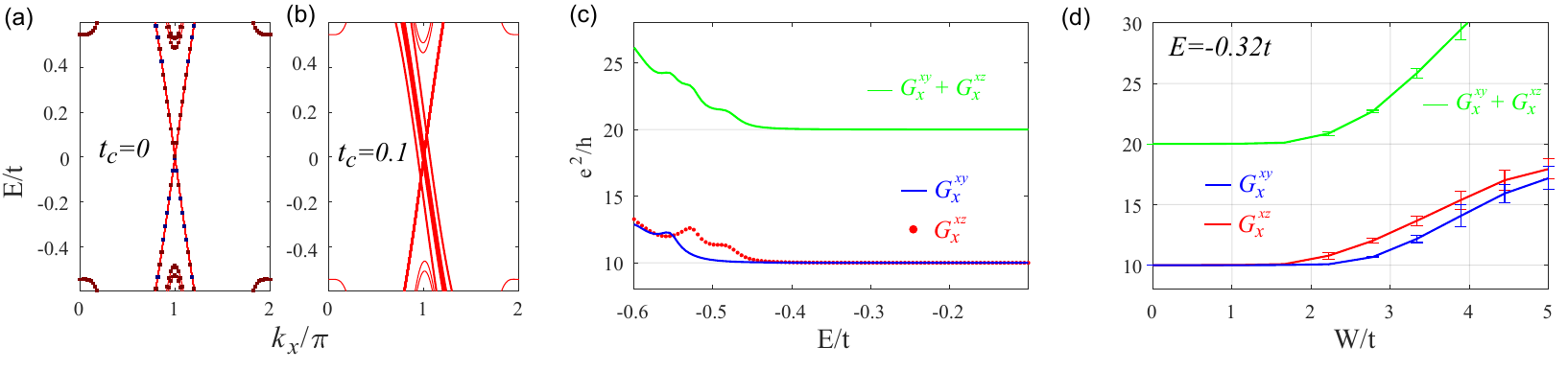}
\caption{(a)-(b) The band structures along the $k_x$. The sample sizes are $L_y=L_z=10$ (c) and (d) The corresponding two-terminal conductances versus Fermi energy $E$ and disorder strength $W$ ,respectively. The sample sizes are fixed at $L_x=L_y=L_z=10$, $N=0$ with $t_c=0.1t$.
 }
\label{S0}
\end{figure*}

\section{III. Details of Bands along the $x$ and $y$ direction}

 \begin{figure*}[t]
\includegraphics[width=16cm]{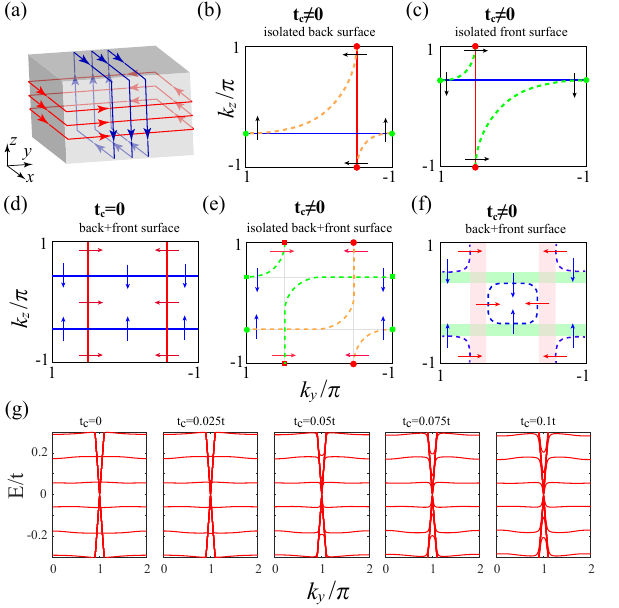}
\caption{(a) The schematic plot of the sample and the chiral surface states. (b)-(f) The schematic plot of the Fermi surfaces of the chiral surface states in the front and back ($y$-$z$) surfaces under different conditions. The blue and red solid lines are the Fermi surfaces for the red and blue chiral modes with $t_c=0$. The arrows roughly mark the group velocity of the states. (g) The band structures $E$ versus $k_y$ for different $t_c$.
 }
\label{S6}
\end{figure*}

Firstly, we illuminate that the coupling $t_c$ has negligible influences to the transport properties and the band structures for a ribbon along the $x$ direction. As shown in Figs. \ref{S0}(a)-(b), The surface states' chirality holds for $t_c\neq0$. Generally, it is attributed to the absent of the networks for the $x$-$y$ and $x$-$z$ surfaces states. The quantized transport properties are robust against disorder and are independent of $t_c$, as plotted in Figs. \ref{S0}(c)-(d).

As a comparison, the bands along the $y$ direction are obviously $t_c$-dependent. As shown in Fig. \ref{S6}(g), the backscattering free chiral modes can be gapped out by turning on $t_c$, and the gaps are determined by $t_c$. These features are consistent with the transport properties in the main text. Generally, the gapped red chiral modes along $k_y$ can be understood as follows:

  (i) For $t_c=0$, the red and blue chiral modes are decoupled, where their Fermi surfaces satisfy the plots in Fig. \ref{S6}(d).

  (ii) After turning on $t_c\neq0$, the Fermi surfaces of the isolated front and back surfaces varies from the solid lines to the dashed lines as plotted in Figs. \ref{S6}(b)-(c). Nevertheless, the dashed and solid lines are both topological nontrivial in the Brillouin zone. Since they both can not continuous deformed into a point, the dashed and solid lines both traversal the entire $k_{y/z}\in[0,2\pi]$.
It indicate that the number of chiral modes along the $y$ direction is $t_c$ independent for the isolated $y$-$z$ surfaces. In other words, the chiral modes can not be gapped out for the isolated $y$-$z$ surfaces.

  (iii)
 In real systems, the required isolated $y$-$z$ surfaces are not available for the considered sample. For example, the chiral turning on the top/bottom surfaces leads to an effective coupling between the front and back surfaces. Putting the two Fermi-surfaces together (see Fig. \ref{S6}(e)) and considering their effective coupling, the intersection of the two Fermi surfaces should be released. Then, the Fermi surfaces in Fig. \ref{S6}(e) will modified into Fig. \ref{S6}(f), where a closed and topologically trivial Fermi surfaces are obtained.

  (vi) Considering the colored regions between the Fermi surfaces, these momentums are unavailable for the chiral modes. Thus, for a particular Fermi energy $E$, the number of chiral modes along $k_y$ should decrease accordingly.

  (v)  By increasing $t_c$, the colored regions should increase. Therefore, the number of chiral modes decreases by increasing $t_c$. These features are consistent with the band structures in Fig. \ref{S6}(g).

\section{IV. Robust and quantized $G^{(xz)}_x\!+\!G^{(xy)}_y$}

We present more details of the robustness and the quantized $G^{(xz)}_x\!+\!G^{(xy)}_y$
In theory, the quantized response $G^{(xz)}_x\!+\!G^{(xy)}_y=(L_y\!+\!L_z)e^2/h$ can be understood as follows.

\begin{figure*}[h]
\includegraphics[width=17cm]{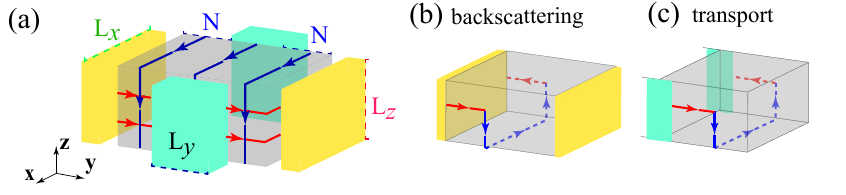}
\caption{(a) Schematic distribution of the chiral surface states with Chern vector $\textbf{C}=(0,1,1)$. The sample sizes has been marked in the figure. (b) and (c) present the influences of chiral networks for different leads.
 }
\label{transportscheme}
\end{figure*}

First of all, when $G^{(xy)}_y$ deviates from its ideal value $L_ze^2/h$, there are $(L_ze^2/h\!-\!G^{(xy)}_y)/(e^2/h)$ red backscattering channels follow the localized trajectory in Fig. \ref{transportscheme}(b).
Significantly, the localized trajectory is available when the chiral surfaces states on both front, back and bottom surfaces are involved.
Thus, these trajectories will contribute the transport for surface conductance $G^{(xz)}_x$, as plotted in Fig. \ref{transportscheme}(c).

Notably, the contribution of these trajectories to $G^{(xz)}_x$ can be quantitatively determined.
Due to the chirality of these states, their transport properties can be simplified as a standard channel matching problem between the front/back and bottom surfaces in mesoscopic transport \cite{LB1}. In specific, the $N\rightarrow \infty$ blue chiral channels on the bottom surface perfectly match with the $(L_ze^2/h-G^{(xy)}_y)h/e^2$ red chiral channels on the front and back surfaces.
Thus, the contribution of these backscattering routes to $G^{(xz)}_x$ equals $(L_ze^2/h-G^{(xy)}_y)$. While $G^{(xy)}_y$ decreases by $(L_ze^2/h-G^{(xy)}_y)$, $G^{(xz)}_x$ will increase by $(L_ze^2/h-G^{(xy)}_y)$ accordingly, which preserves the relation:
  \begin{equation}
  G^{(xz)}_x\!+\!G^{(xy)}_y\!=\![L_ye^2/h\!+\!(L_ze^2/h\!-\!G^{(xy)}_y)]\!+\!G^{(xy)}_y\!=\!(L_y\!+\!L_z)e^2/h.
  \end{equation}
Here, $L_ye^2/h$ is the intrinsic contribution of $G_x^{(xz)}$, which already exists for $t_c=0$.

The disorder only tend to enhance the scattering between the chiral networks, which simply modifies the effective coupling $t_c$.

\section{V. Kirchhoff's equation of Hall bar setups}

\subsection{A. current along $y$ direction}

 \begin{figure*}[h]
\includegraphics[width=17cm]{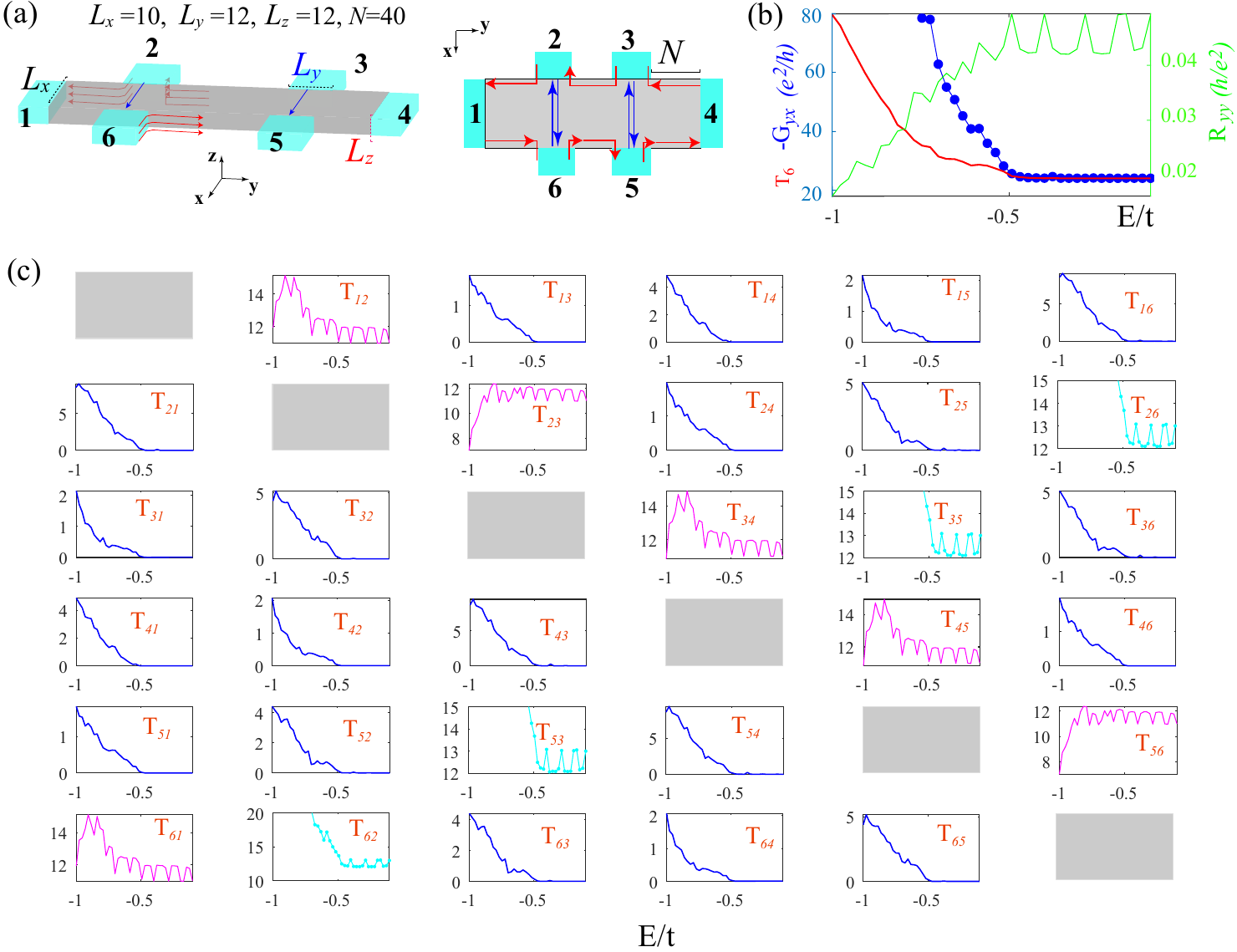}
\caption{(a) The schematic plot of the Hall bar setup. The red and blue solid line with arrows represent the chiral surface states. The definition of sample sizes has been marked in the figure. The distances between different leads are set as $N=40$.
(b) The Hall $G_{yx}=I/(V_2-V_6)$ (blue solid line), $T_6=\sum_{j\neq6}T_{6,j}$ (red solid line) and longitudinal resistance $(V_2-V_3)/I$ (green solid line).
(c) The transmission coefficients between different leads. The pink solid lines roughly present $T_{j,j+1}=T_{61}\sim G^{(xy)}_y$. The cyan solid lines roughly present  $T_{6 2}\approx G_x^{(xz)}$. The extra transmission coefficients can be neglected for $E\in(-0.5t,-0.1t)$. In our calculation, we set $t_c=0.1t$.
 }
\label{S2}
\end{figure*}

In this section, we present the Kirchhoff's equation $\textbf{I}_y=\textbf{T}\textbf{V}$ of the Hall bar setup in Fig. \ref{S2}(a). The sample sizes has been marked in the figure. For the six-terminal devices, the  Kirchhoff's equation $\textbf{I}_y=\textbf{T}\textbf{V}$ can be rewritten as \cite{LB1}:
\begin{equation}
\left(
\begin{array}{cccccc}
~T_{1} & -T_{12} & -T_{13} & -T_{14} & -T_{15} & -T_{16}\\
-T_{21} & ~T_{2} & -T_{23} & -T_{24} & -T_{25} & -T_{26}\\
-T_{31} & -T_{32} & ~T_{3} & -T_{34} & -T_{35} & -T_{36}\\
-T_{41} & -T_{42} & -T_{43} & ~T_{4} & -T_{45} & -T_{46}\\
-T_{51} & -T_{52} & -T_{53} & -T_{54} & ~T_{5} & -T_{56}\\
-T_{61} & -T_{62} & -T_{63} & -T_{64} & -T_{65} & ~T_{6}
\end{array}
\right)
\left(
\begin{array}{c}
~~V\\
~~V_2\\
~~V_3\\
-V\\
~~V_5\\
~~V_6
\end{array}
\right)
=
\left(
\begin{array}{c}
~~I_y\\
~~0\\
~~0\\
-I_y\\
~~0\\
~~0
\end{array}
\right).
\end{equation}
$V$ and $-V$ are the voltage potentials on the lead-1 and lead-4. $V_n$ are the voltage potentials on the lead-$n$ with $n\in[2,3,5,6]$.
$I_{y}=I$ is the current between the lead-1 and lead-4 along the $y$ direction.
Here, one sets $T_n=\sum_{m\neq n}T_{n,m}$ with $T_{nm}={\rm Tr}[\Gamma_nG^r\Gamma_mG^a]$ and $n,m\in[1,2,3,4,5,6]$. The retarded Green's function satisfies $G^r=[E+i\eta-\mathcal{H}-\sum_{n\in[1-6]}\Sigma_n]^{-1}$ for the six-terminal systems. $\Sigma_n$ is the same as those in the two-terminal cases.

Noticing the transmission coefficient in Fig. \ref{S2}(c), the Kirchhoff's equation for the quantized energy region with $E\in[-0.5t,-0.1t]$ can be simplified as:
\begin{equation}
\left(
\begin{array}{cccccc}
~T_{1} & -T_{12} & 0 & 0 & 0 & 0\\
0 & ~T_{2} & -T_{23} & 0 & 0 & -T_{26}\\
0 & 0 & ~T_{3} & -T_{34} & -T_{35} & 0\\
0 & 0 & 0 & ~T_{4} & -T_{45} & 0\\
0 & 0 & -T_{53} & 0 & ~T_{5} & -T_{56}\\
-T_{61} & -T_{62} & 0 & 0 & 0 & ~T_{6}
\end{array}
\right)
\left(
\begin{array}{c}
~~V\\
~~V_2\\
~~V_3\\
-V\\
~~V_5\\
~~V_6
\end{array}
\right)
=
\left(
\begin{array}{c}
~~I_\alpha\\
~~0\\
~~0\\
-I_\alpha\\
~~0\\
~~0
\end{array}
\right).
\label{KE1}
\end{equation}
Transmission coefficients marked in blue solid lines are set as zero since they are much smaller than the transmission coefficients marked in pink and cyan solid lines in Fig. \ref{S2}(c).

Based on the chirality of the surface states, one should also notice the following relations should holds for the Chern vector protected quantized Hall conductances:
\begin{equation}
\begin{array}{l}
T_{62}= T_{26}= T_{35}= T_{53}\approx G_{x}^{(xz)}h/e^2,\\
T_{61}= T_{12}= T_{23}= T_{34}= T_{45}=T_{56}\approx G_{y}^{(xy)}h/e^2.
\end{array}
\end{equation}
Here, $G_{x}^{xy}$ ($G_{y}^{yz}$) is the two terminal conductance along $x$ ($y$) directions contributed by the chiral surface states located on the $x$-$z$ ($x$-$y$) boundaries.
Considering $G_{x}^{(xz)}+G_{y}^{(xy)}=(mL_y+nL_z)e^2/h$, it illuminates the quantized $T_6=\sum_{j\neq6}T_{6,j}\approx T_{61}+T_{62}=mL_y+nL_z$, as shown in Fig. \ref{S2}(b). Considering the restriction of the transmission coefficients, equation (\ref{KE1}) gives rise to
\begin{equation}
G_{yx}=I/(V_2-V_6)=-(T_{61}+T_{62})e^2/h=-[G_{x}^{(xz)}+G_{y}^{(xy)}]=-(mL_y+nL_z)e^2/h.
\end{equation}

\subsection{B. current along $x$ direction}

 \begin{figure*}[h]
\includegraphics[width=17cm]{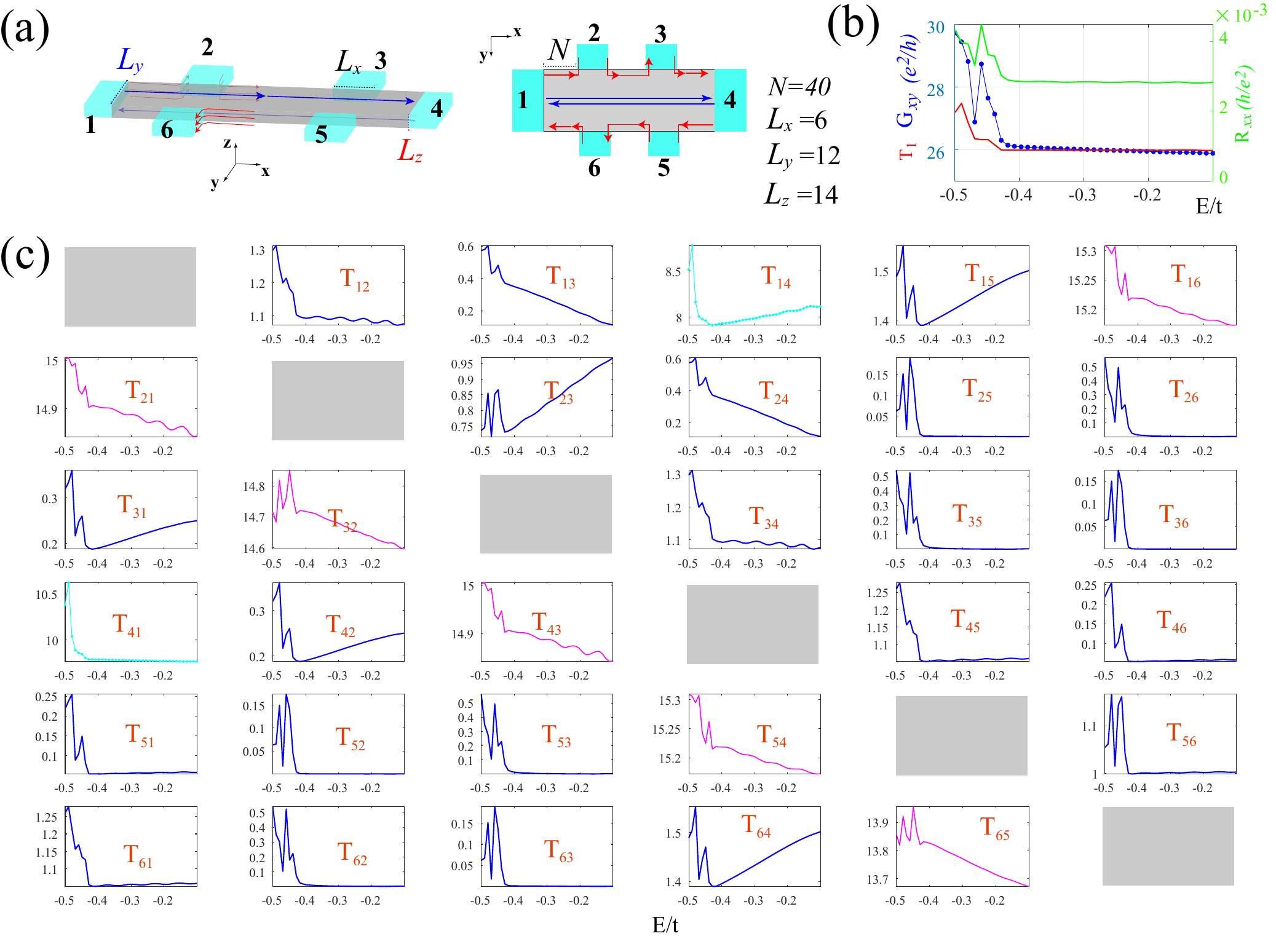}
\caption{(a) The schematic plot of the Hall bar setup. The red and blue solid line with arrows represent the chiral surface states. The definition of sample sizes has been marked in the figure. The distances between different leads are set as $N=40$.
(b) The Hall $G_{xy}=I/(V_2-V_6)$ (blue solid line), $T_1=\sum_{j\neq6}T_{1,j}$ (red solid line) and longitudinal resistance $(V_2-V_3)/I$ (green solid line).
(c) The transmission coefficients between different leads. The pink solid lines roughly present $T_{j+1,j}=T_{16}\sim G^{(xy)}_x$. The cyan solid lines roughly present  $T_{14}\sim G_x^{(xz)}$. The extra transmission coefficients can be neglected for $E\in[-0.4t,-0.1t]$. In our calculation, we set $t_c=0.1t$.
 }
\label{S3}
\end{figure*}

Generally, similar analysis for currents along the $x$ direction as shown in Fig. \ref{S3}. The Kirchhoff's equation $\textbf{I}_x=\textbf{T}\textbf{V}$ of Hall bar setup in Fig. \ref{S3}(a) can be roughly simplified as:
\begin{equation}
\left(
\begin{array}{cccccc}
~T_{1} & 0 & 0 & -T_{14} & 0 & -T_{16}\\
-T_{21} & ~T_{2} & 0 & 0 & 0 & 0\\
0 & -T_{32} & ~T_{3} & 0 & 0 & 0\\
-T_{41} & 0 & -T_{43} & ~T_{4} & 0 & 0\\
0 & 0 & 0 & -T_{54} & ~T_{5} & 0\\
0 & 0 & 0 & 0 & -T_{65} & ~T_{6}
\end{array}
\right)
\left(
\begin{array}{c}
~~V\\
~~V_2\\
~~V_3\\
-V\\
~~V_5\\
~~V_6
\end{array}
\right)
=
\left(
\begin{array}{c}
~~I_x\\
~~0\\
~~0\\
-I_x\\
~~0\\
~~0
\end{array}
\right).
\label{KE2}
\end{equation}
Transmission coefficients marked in blue solid lines are set as zero since they are much smaller than the transmission coefficients marked in pink and cyan solid lines in Fig. \ref{S3}(c) while sample size $L_y\rightarrow\infty$ and $L_z\rightarrow\infty$.

Based on the chirality of the surface states, one should also notice the following relations should holds for the Chern vector protected quantized Hall conductances:
\begin{equation}
\begin{array}{l}
T_{14}= T_{41}\approx G_{x}^{(xz)}h/e^2,\\
T_{16}= T_{21}= T_{32}= T_{43}= T_{54}=T_{65}\approx G_{x}^{(xy)}h/e^2.
\end{array}
\end{equation}
Here, $G_{x}^{(xz)}$ ($G_{x}^{(xy)}$) is the two terminal conductance along $x$ ($x$) directions contributed by the surface states located on the $x$-$y$ ($x$-$z$) boundaries.
Considering $G_{x}^{(xy)}+G_{x}^{(xz)}=mL_y+nL_z$, it illuminate the quantized $T_1=\sum_{j\neq1}T_{1,j}\approx T_{16}+T_{14}=mL_y+nL_z$, as shown in Fig. \ref{S3}(b). Considering the restriction of the transmission coefficients, equation (\ref{KE2}) give rise to the quantized
\begin{equation}
G_{xy}=I/(V_2-V_6)=(T_{16}+T_{14})e^2/h=G_{x}^{(xy)}+G_{x}^{(xz)}=(mL_y+nL_z)e^2/h.
\end{equation}

\section{VI. Disorder effects on the quantized Hall conductance}

 \begin{figure*}[h]
\includegraphics[width=16cm]{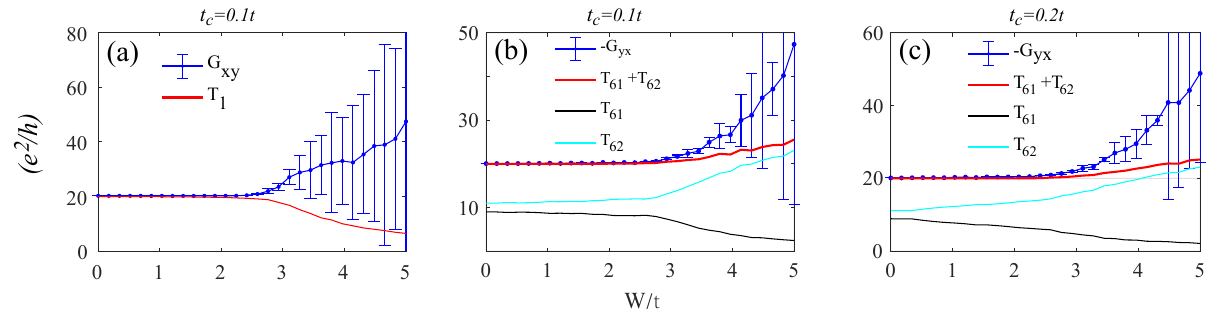}
\caption{(a) and (b) The Hall conductance (blue solid line) and the response function (blue solid line) versus disorder strength $W$. The sample sizes are $L_x=L_y=L_z=10$, $N=40$ and $E=-0.2t$.}
\label{S4}
\end{figure*}

Fig. \ref{S4} shows the quantized Hall conductance versus the disorder strength \cite{Anderson0}. The Anderson disorder $H_w=\varepsilon_n c^\dagger_nc_n$ with $\varepsilon_n\in[-W/2,W/2]$ is considered, and $W$ is the disorder strength.
The Chern vector protected quantized Hall conductance and the proposed relationships are robust against disorder, as plotted in Figs. \ref{S4}(a)-(c). Further, the quantized $G_{yx}$ is independent of the quantization of $T_{61}$ or $T_{62}$. $T_{61}$ and $T_{62}$ for the Hall bar setups could deviate from their quantized values with $T_{61}+T_{62}$ still preserves its quantized value. These features are consistent with the proposed scheme for $G_{yx}$.